\documentclass{webofc}
\usepackage[varg]{txfonts}   
\usepackage{changes}
\usepackage{amsmath,graphicx,multirow,tabularx,physics}
\usepackage{davitex,slashed,listings}

\newcommand{\muCEP}{\mu_B^{\text{CEP}}}
\newcommand{\TCEP}{T^{\text{CEP}}}
\newcommand{\zLYE}{z_{\rm LYE}}
\newcommand{\muLYE}{\mu_{\text{LYE}}}
\begin{document}

\title{Exploring the Critical Points in QCD with Multi-Point Padé and Machine Learning Techniques in (2+1)-flavor QCD}

\author{\firstname{Jishnu} \lastname{Goswami}\inst{1}\fnsep\thanks{\email{jishnu.goswami@riken.jp}} \and
        \firstname{D. A.} \lastname{Clarke}\inst{2}\and
        \firstname{P.} \lastname{Dimopoulos}\inst{2}\and
        \firstname{F.} \lastname{Di Renzo}\inst{3}\and
        \firstname{C.} \lastname{Schmidt}\inst{4}\and
        \firstname{S.} \lastname{Singh}\inst{4}\and
        \firstname{K.} \lastname{Zambello}\inst{5}
}

\institute{RIKEN Center for Computational Science, Kobe 650-0047, Japan
\and
Department of Physics and Astronomy, University of Utah, Salt Lake City, Utah 84112, United States 
\and 
Dipartimento di Scienze Matematiche, Fisiche e Informatiche, Università di Parma and INFN, Gruppo Collegato di Parma
I-43100 Parma, Italy
\and
           Fakult\"at f\"ur Physik, Universit\"at Bielefeld, D-33615 Bielefeld, Germany
\and
Dipartimento di Fisica dell’Università di Pisa and INFN--Sezione di Pisa, 
Largo Pontecorvo 3, I-56127 Pisa, Italy.
          }


\abstract{
Using simulations at multiple imaginary chemical potentials for $(2+1)$-flavor QCD, we construct multi-point Padé approximants. We determine the singularties of the Padé approximants and demonstrate that they are consistent with the expected universal scaling behaviour of the Lee-Yang edge singularities. We also use a machine learning model, Masked Autoregressive Density Estimator (MADE), to estimate the density of the Lee-Yang edge singularities at each temperature. This ML model allows us to interpolate between the temperatures. Finally, we extrapolate to the QCD critical point using an appropriate scaling ansatz.}

\maketitle

\section{Introduction}\label{sec:intro}
Our ability to predict thermodynamic observables and determine the QCD critical point at real values of chemical potentials is severely limited by the infamous sign problem. To address this issue, there are two common approaches: expanding the QCD partition function in a Taylor series with respect to the charge chemical potentials ($\mu_B,\mu_Q,\mu_S$) \cite{Gavai:2001fr,Allton:2002zi} or analytically continuing from imaginary chemical potentials \cite{Borsanyi:2012cr,Guenther:2017hnx}. However, both methods have limitations, particularly for larger values of the baryon chemical potential. Recently, we proposed a multi-point Padé approach~\cite{Nicotra:2021ijp,Dimopoulos:2021vrk}.

In this proceeding, we will focus on the QCD critical point, which will emerge at a real value of the baryon chemical potential. We use the multi-point Padé approach to locate the Lee-Yang edge (LYE) singularities in the complex chemical potential plane, which are obtained from lattice QCD simulated data. The universal scaling of singularities in the vicinity of the QCD critical endpoint is also investigated. LYEs associated with the QCD critical point at real chemical potential for various temperatures are calculated. As the temperature decreases, the imaginary part of the singularities becomes smaller, hinting at the possible existence of a critical point at low temperature. A machine learning technique is used to model the probability density of the singularities and interpolate the real and imaginary parts of the singularities between different temperatures. By employing a suitable scaling ansatz, the singularities are extrapolated towards the real axis, and the possible location of the QCD critical point is estimated. Preliminary results of ($T_c^{CEP},\mu_B^{CEP}$) are consistent with model predictions and other lattice QCD calculations~\cite{Bollweg:2021vqf,Goswami:2022nuu,Clarke:2022pfz}. Recently, there are also other estimates for the QCD critical point from LYEs~\cite{Basar:2021gyi,Basar:2023nkp}.

In the following sections we present the LYEs obtained for (2+1)-flavor QCD at each temperature. We use two methods, one based on a universal scaling ansatz and one based on a machine learning ansatz, to interpolate the data between the temperatures. Finally, we extrapolate the LYEs to the QCD critical point. We performed simulations at four temperature values for the lattice size $36^3\times6$ with multiple imaginary chemical potential values. The temperature scales are chosen according to the parametrization given in~\cite{Bollweg:2021vqf} in the $f_K$ scale.
\section{LYEs and related scaling of the the QCD critical point}
The universal scaling function in the vicinity of a second-order phase transition can be written
\begin{equation}\begin{aligned}
f(t,h)
&\simeq  b^{-d}f_s(b^{1/\nu} t/t_0, b^{\beta\delta/\nu} 
h/h_0) \nonumber + \rm{regular} \  \rm{terms} \; \\
f(z)&\simeq h^{\frac{2-\alpha}{\beta}}f_s(z) + \ \rm{regular~terms}, 
\label{scalingfct}
\end{aligned}\end{equation}
where $z= t/h^{1/\beta\delta}$ and $f_s$ is the universal scaling function of free energy of the
temperature-like scaling field $t$
and magnetization-like scaling field $h$. The temperature-like scaling field is related to the reduced temperature direction and couples to the energy-like observable. Meanwhile, $h$ is related to the 
symmetry-breaking direction and couples to the magnetization-like observable. 
Here $\beta$, $\gamma$, $\delta$ and $\nu$
are critical exponents and $t_0$ and $h_0$
are non-universal scale parameters.
We use the following scaling relations to map $T$ and $\mu_B$ to 
the temperature-like and magnetization-like scaling directions $t$ and $h$:
\begin{figure}
    \centering
    \begin{minipage}[b]{0.4\linewidth}
        \includegraphics[width=\linewidth]{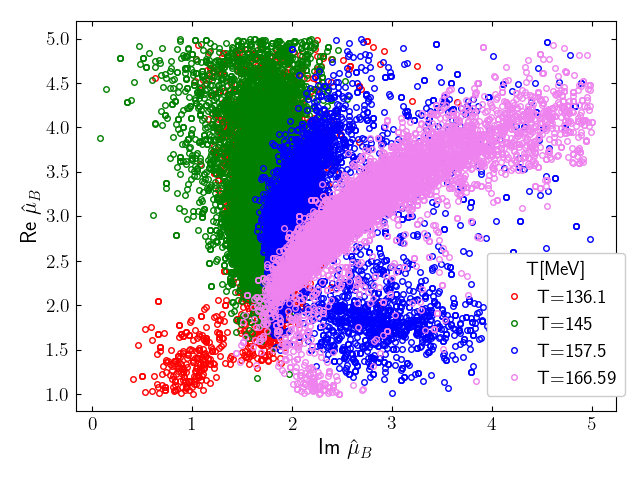}
        \caption{Distribution of the poles at different temperature for $36^3\times6$ lattice.}
        \label{fig:first_figure}
    \end{minipage}
    \hfill
    \begin{minipage}[b]{0.4\linewidth}
        \includegraphics[width=\linewidth]{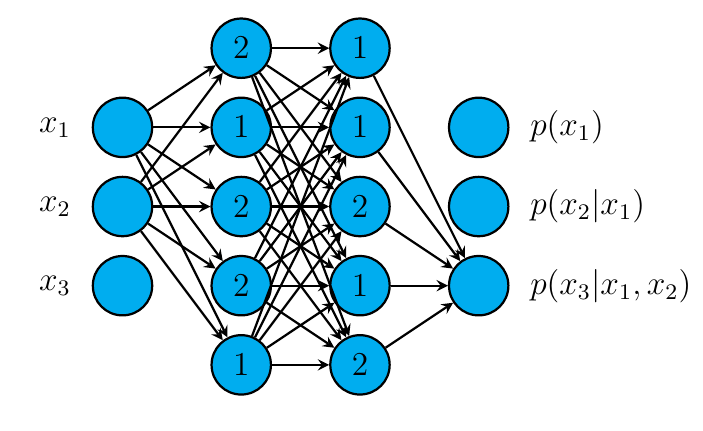}
        \caption{A schematic diagram of the MADE block.}
        \label{fig:second_figure}
    \end{minipage}
\end{figure}

\begin{equation}\begin{aligned}
t &=A_t\Delta T+B_t \Delta\mu_B,\\
h &=A_h\Delta T+B_h \Delta\mu_B, 
\label{eq:CEPansatz}
\end{aligned}\end{equation}
with $\Delta T\equiv T-\TCEP$, $\Delta \mu_B\equiv\mu_B-\muCEP$, and
constants $A_i$, $B_i$, $i=t,h$.
We would like to follow the path of the LYE expressed
in terms of the scaling variable $z$.
This leads to
\begin{equation}\label{eq:zLYE}
    \zLYE = |\zLYE|~e^{i\pi/2\beta\delta}.
\end{equation}
Plugging \equatref{eq:CEPansatz} into \equatref{eq:zLYE} implies
that $\muLYE$ should scale~\cite{Stephanov:2006dn} as
\begin{equation}\begin{aligned}
      \Re\muLYE&=\muCEP+ c_{10}\Delta T +  c_{11}\Delta T^2 + O(\Delta T^3)  \\
      \Im\muLYE&=ic_{20}|z_c|^{-\beta\delta}\Delta T^{\beta\delta}.
\end{aligned}\end{equation}


\section{Machine learning model for the LYEs and Extrapolation of the LYEs to the critical point}\label{sec:setup}
\begin{figure}[ht]
    \centering
    \begin{minipage}[b]{0.4\linewidth}
        \includegraphics[width=0.85\linewidth]{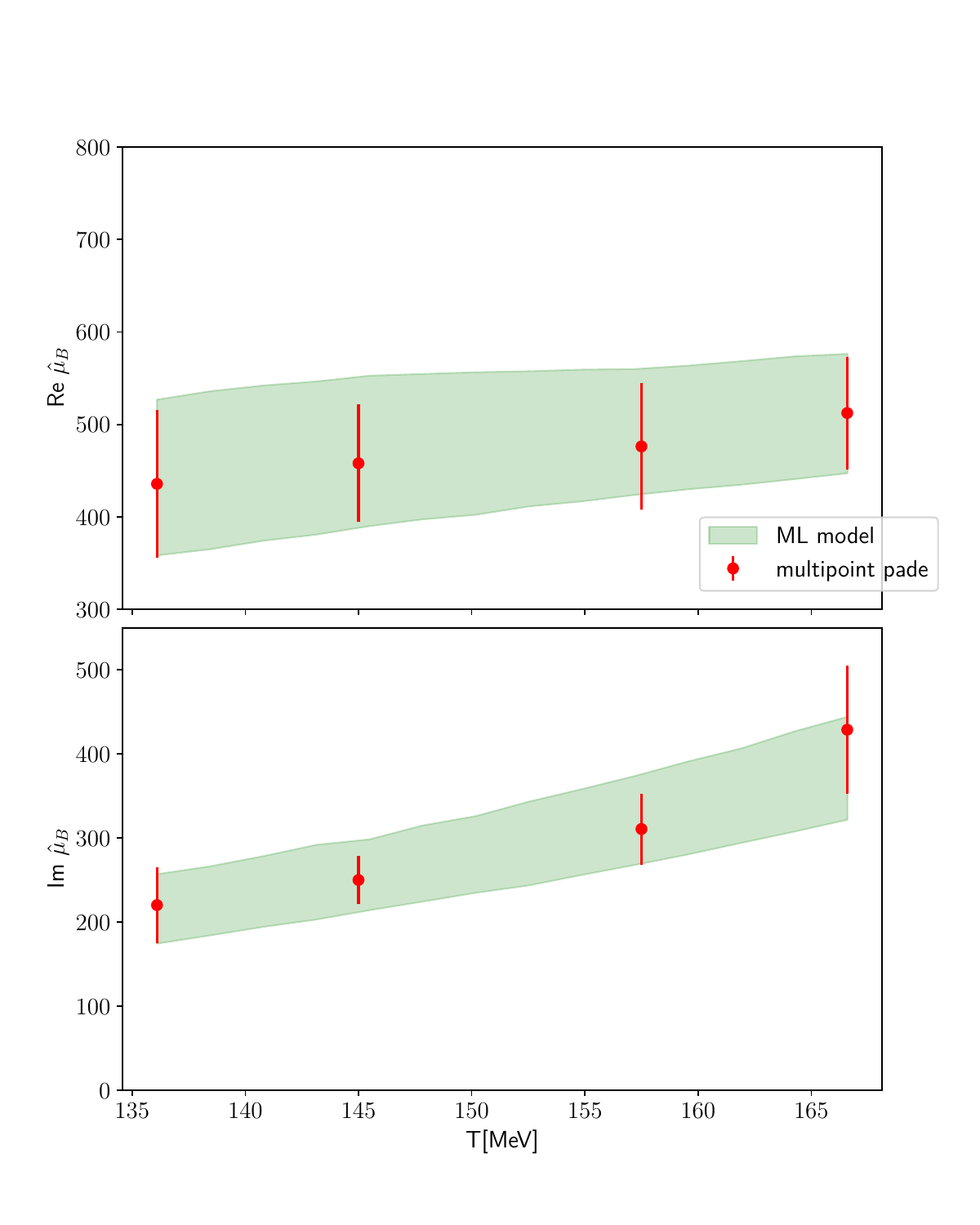}
        \caption{Interpolation of the LYEs, based on ML model.}
        \label{fig:mlmodel}
    \end{minipage}
    \hfill
    \begin{minipage}[b]{0.4\linewidth}
        \includegraphics[width=\linewidth]{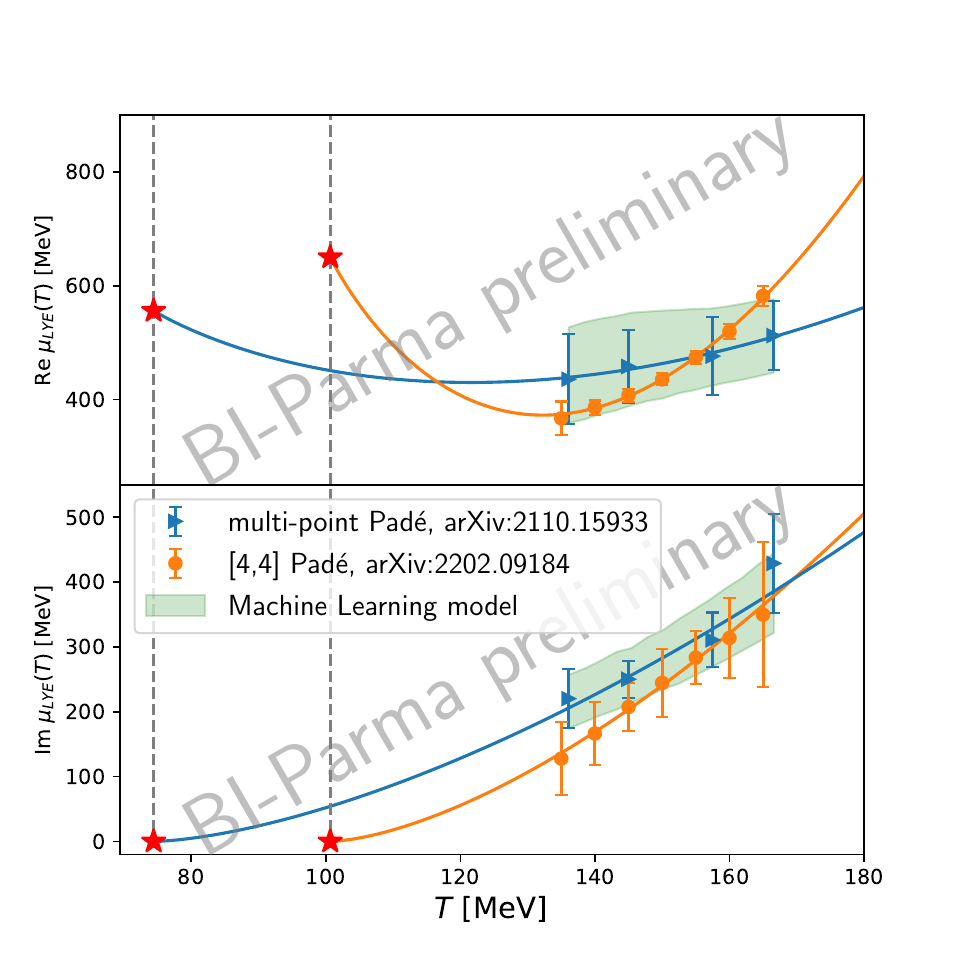}
        \caption{Extrapolation of the LYEs to the QCD critical point. We also compare them with the singularity obtained in \cite{Bollweg:2022fqq}}
        \label{fig:CEPextra}
    \end{minipage}
\end{figure}

We generate scatter plots of the densities of the LYEs via bootstrap sampling methods on data sets, shown in \figref{fig:first_figure}. This study primarily targets the probability density functions of the LYEs at specific temperatures. The methodology involves employing a Masked Autoregressive Flow (MAF), which is composed of several Masked Autoregressive Density Estimator (MADE)~\cite{germain2015made,papamakarios2018masked} segments. These segments are embedded within a fully-connected network architecture.
The masking technique is individually applied to each layer of the network, as illustrated in \figref{fig:second_figure} and outlined in MADE's Python implementation~\cite{germain2015made,papamakarios2018masked}. Moreover, a MADE block effectively breaks down a multivariate probability distribution into a sequence of conditional univariate distributions. Remarkably, MAF has been applied to ascertain the critical mass in the study of degenerate 5-flavor QCD, as shown in \cite{Karsch:2022yka}. In this study, we use the $\Re {\mu}_B$ and $\Im {\mu}_B$ of the LYEs shown in \figref{fig:first_figure} to train the model and estimate the probability distribution $p(\rm{Re} \  \hat{\mu}_B, Im \ \  \hat{\mu}_B| T)$. This can be further extended to study the chiral LYEs as well, $p(\rm{Re} \  \hat{\mu}_B, Im \ \  \hat{\mu}_B| T, m_l, N_{\sigma})$.
In \figref{fig:mlmodel} we show the results from the machine learning model and perform an extrapolation to the critical point in \figref{fig:CEPextra}.

\section{Summary and outlook}\label{sec:outlook}
In this study, we have explored the QCD critical point using techniques involving multi-point Padé approximants and machine learning in (2+1)-flavor QCD. 
We constructed multi-point Padé approximants from simulations at various imaginary chemical potentials, which enabled us to determine the singularities of these approximants. These singularities were found to be consistent with the expected Lee-Yang edge singularities, validating our approach. Furthermore, the application of the Masked Autoregressive Density Estimator (MADE) model for estimating the density of LYEs across different temperatures has been a novel aspect of our study. This machine learning technique allowed us to interpolate between temperatures and furthermore extrapolate our findings to predict the QCD critical point's location.
In the future, we are also aiming to determine the LYEs related to chiral phase transition of QCD. And, our machine learning model can be also of great use.
\section{Acknowledgements}
This work is supported by the Deutsche Forschungsgemeinschaft (DFG, German Research Foundation) through grant number 315477589, from the European Union under grant agreement No. H2020-MSCAITN-2018-813942, and the I.N.F.N. under the research project i.e. QCDLAT. DAC was supported by the National Science Foundation under Grants PHY20-13064 and PHY23-10571.
\bibliography{bibliography.bib}

\end{document}